# Scaling conditional tail probability and quantile estimators

## JOHN COTTER[a]

[a] Centre for Financial Markets, Smurfit School of Business, University College Dublin, Carysfort Avenue, Blackrock, Co. Dublin, Ireland. Email: john.cotter@ucd.ie.

This study was carried out while Cotter was visiting the UCLA Anderson School of Management and is thankful for their hospitality. Cotter's contribution to the study has been supported by a University College Dublin School of Business research grant. The author would like to thank the Editor and two anonymous referees for helpful comments.

March 2009

*We present a novel procedure for scaling relatively high frequency tail probability and quantile estimates for the conditional distribution of returns.*

**Introduction**

A key issue for risk management in practice is to decide the relevant horizon associated with risk measurement. Many different horizons may be relevant from short (eg. daily) to long (eg. monthly) timeframes and the risk manager must be able to provide measures across a range of horizons.[1] This article measures risk at different horizons using volatility forecasts at high frequency as inputs that are then scaled for longer horizons.

In terms of risk measurement probability and quantile risk estimation has developed enormously in the past decade from Value at Risk (VaR) measures to coherent measures such as Expected Shortfall. These measures allow the investor to determine their risk profile accounting for losses (quantiles) at a given likelihood (probability) and a given time frame (holding period).

Within risk estimation two key modelling features attracting enormous attention throughout time are the fat-tailed (eg. see Mandelbrot, 1963) and volatility clustering (eg. see Bollerslev, 1986) properties inherent in financial data. These features exist for different holding periods albeit in an inconsistent manner.

Here we present a framework that addresses these features and allows for applying a simple scaling rule for the risk measures across different (eg. lower frequency) holding periods. We refer to this framework as conditional EVT. To begin we use an AR (1)-GARCH (1, 1) specification with student-t innovations to model the conditional distribution (such models have attracted considerable success in forecasting volatility for short horizons). Our choice of the specific GARCH model follows McNeil and Frey (2000) and is similar in objective to the use of the ARMA-GARCH model by Barone-Adesi et al (1999). First by fitting this model to returns we obtain forecasts of the conditional mean through the AR component of the filter and the conditional variance through the GARCH component. We choose the commonly applied t-distribution to model fat-tail innovations from a range of candidate distributions (eg. stable distribution). Second, our GARCH conditional volatility series is also used to filter the returns series resulting in identical and independently distributed (iid) filtered returns.

We then apply Extreme Value Theory (EVT) to the conditional filtered series to model the tail returns allowing us to examine low probability (out-of-sample) events for single-period horizons. These estimates are scaled with the EVT $\alpha$-root scaling law that require iid realisations to give risk measures for multi-period horizons.[2] We illustrate our modelling approach with a simulation study and an application to daily S&P500 index returns.

**Why use conditional risk measures?**
Conditional risk measures are important as investors have an interest in obtaining risk measures from the conditional return distribution as part of their risk management strategy. The conditional measures provide time-varying risk estimates updated by current volatility dynamics thereby allowing the investor manage the ongoing and changing risk of their investment as the distribution of returns changes over time. Investors are interested in obtaining these risk estimates for different frequencies that correspond to potential holding periods (for instance they may need to meet regulatory requirements such as a 10-day VaR or support trading activity by having a 1-day VaR).

---

[1] For instance Christoffersen and Diebold (2000) note that the relevant horizon will vary by position (eg. trading desk vs. Chief Financial Officer), by motivation (eg. private vs. regulatory) and by other concerns such as industry type (banking vs. insurance) etc.
[2] A referee has quite rightly pointed out that scaling GARCH volatility estimates may result in an estimation problem if volatility was to deviate substantially from current levels. However, other alternative approaches such as using average volatility or assuming mean reversion results in the same potential estimation problem. The approach followed here does allow the risk manager to have multi-period risk forecasts based on current GARCH volatility that have been found to model time-varying

We benchmark our modelling approach of conditional EVT risk measures (note we are not suggesting that we scale the GARCH volatility estimates directly) against those obtained from the thin-tailed Gaussian distribution. The Gaussian benchmark is chosen as it is a commonly applied model for financial time series (eg. RiskMetrics VaR measures assume conditional normality) but does not account for fat-tails. Both approaches have a number of similarities such as having a scaling rule for obtaining multi-period risk estimates from single-period estimates and allowing for extrapolation to out-of-sample probability levels. In an unconditional setting, EVT has been found to dominate Gaussian (and other) measures in modelling tail risk as it (through the Fréchet distribution) is more accurate in dealing with fat-tails (Danielsson and de Vries, 2000) and results in increased accuracy as you move to lower probability events. Specifically, a Gaussian distribution results in underestimated (overestimated) unconditional risk estimates for single-period (multi-period) settings.

Our fitting of the GARCH process results in iid filtered returns, and predictors of conditional returns and volatility through iteration. We then apply the EVT $\alpha$-root scaling law that only requires the existence of a finite variance and an iid series. The scaling procedure advantageously requires no further estimation of any additional parameters and obtains efficiency in the scaling operation by using the highest frequency realisations. Moreover from a modelling procedure, tail estimation for low frequency observations is associated with small sample bias and the alternative of scaling high frequency to low frequency tail estimates reduces the bias.

Scaling is important as it allows one to overcome the lack of non-overlapping returns for low frequency horizons. A number of previous studies have examined scaling.

---

volatility adequately. More important it is the risk measures of the conditional distribution that are scaled after they are estimated using EVT.

Most common has been the use of the square root rule assuming a Gaussian distribution and this has been found to underestimate volatility as you move to longer horizons (Danielsson and Zigrand, 2006). Moreover, Drost and Nijman (1993) demonstrate scaling for GARCH processes whereas Reiss and Thomas (2007) discuss the use of EVT scaling laws. The approach presented here uses the EVT scaling law after first using a GARCH model to give forecasts of conditional volatility and filtering the returns series resulting in iid residuals. By using a semi-parametric tail estimator it avoids the pitfall of assuming that that the initial modelling process exactly fits the data, as Drost and Nijman do in their aggregation of GARCH processes.

**Risk measures and modelling procedure**

Before we outline our modelling procedure we provide details of our risk measures and the environment facing the investor. We define separate probability, $P_{Q,h,t}$, and quantile, $Q_{P,h,t}$, risk measures for the conditional distribution for any holding period, h, and time period, t. The first measure estimates the probability of exceeding a certain loss quantile whereas the second measure estimates the loss quantile for a given probability level. These risk measures provide risk managers with dynamic risk information on prospective losses occurring in a time-varying fashion for single, h = 1, and multi-period, h > 1, settings.

These conditional risk estimates are based on the assumption that the returns series exhibits fat-tails and volatility clustering in line with financial returns (see figure 1). In the time series plots we see periods of high and low volatility. Also in the QQ plot we see from the conditional distribution the existence of fat-tails with both upper and lower tail values diverging from the corresponding Gaussian values and the divergence increasing the further you move out the tail. Thus an investor's risk management strategy would benefit from using risk measures that incorporate these two properties.

INSERT FIGURE 1 HERE

Turning to the modelling procedure we begin by fitting an AR (1)-GARCH (1, 1) model underpinned by student-t innovations with 4 degrees of freedom to the returns series (our choice of degrees of freedom was based on Hill estimator values for the S&P500 returns series). The approach has two aims: to obtain a description of the conditional distribution for both mean and variance and to obtain iid residuals from standardising returns with the GARCH model. Previous studies (eg. McNeil and Frey, 2000) have followed a similar approach. The choice of student-t innovations recognises the fat-tailed property already outlined. Assume that a sequence of returns, R, is related to the residual series Z, mean returns are modelled with and AR(1) process and volatility is modelled with a GARCH (1, 1) process:

$R_t = \mu_t + \sigma_t Z_t$ (1)

$\mu_t = \phi R_{t-1}$

$\sigma^2_t = \alpha_0 + \alpha_1 R_{t-1}^2 + \beta_1 \sigma^2_{t-1}$

for $\alpha_0, \alpha_1,$ and $\beta > 0$; $0 < \alpha_1 + \beta_1 < 1$ and $\beta$ measures the persistence in volatility.

The conditional mean, $\mu_t$, and variance, $\sigma_t$, parameters of the returns distribution are obtained through iterations of the AR and GARCH components of the model respectively. In practice it is the volatility forecasts that are important as daily expected returns are often assumed to be zero. The model also filters the returns series by the conditional volatility series to obtain an iid residual series, Z.

The main assumption of the GARCH model is that the conditional second moment, $\sigma_t$, has a degree of persistence resulting in volatility clustering. Our conditional risk measures are explicitly adapted for this feature by continuously updating for time varying volatility.

Turning to the application, details of fitting the AR-GARCH (1, 1) model with student-t innovations are given in table 1. We see strong persistence of past volatility indicating the tendency to form volatility clusters over time. Prior to fitting the GARCH model the returns series exhibit serial correlation but the filter works well resulting in iid filtered returns (see series Z) allowing the use of the extreme value scaling law.

INSERT TABLE 1 HERE

Second, we detail our conditional risk measures that generate a set of time varying probability and quantile estimates. The probability measure for a single-period is obtained with:

$$P_{Q,t} = \mu_{t+1} + \sigma_{t+1} P_Q[Z_t] \qquad (2)$$

And the quantile measure is given by:

$$Q_{P,t} = \mu_{t+1} + \sigma_{t+1} Q_P[Z_t] \qquad (3)$$

Where the conditional variables, $\mu_{t+1}$ and $\sigma_{t+1}$, give predictive risk estimates accounting for the conditional mean and volatility environment facing the investor.

We obtain the tail conditional probability and quantile risk estimates from using EVT on the filtered iid series, Z. EVT relies on order statistics where the set of filtered returns $\{Z_1, Z_2,..., Z_n\}$ associated with days 1, 2…, T, are assumed to be iid, and belonging to the true unknown distribution F. We examine the maxima, $M_T$ = Max$\{Z_1, Z_2,..., Z_T\}$, of the iid variables where the asymptotic behaviour of tail values is given by the Fisher-Tippett theorem. This theorem separates out three limiting distributions (Fréchet, Gumbell and Weibull) based on $\alpha$, the tail index, where asymptotic convergence occurs using Gnedendko's theorem. As we have seen financial returns exhibit fat-tails (eg. the kurtosis statistic for our S&P500 data is 12.17) and this corresponds to a fat-tailed extreme value distribution, the Fréchet distribution. The tail of a Fréchet distribution (eg. Student-t) has a power decline where not all moments are necessarily defined. In contrast, the tail declines exponentially for a Gumbell distribution (eg. Gaussian) with all moments defined, and no tail is defined for a Weibull distribution (eg. Uniform) beyond a certain threshold. For the Fréchet distribution Gnedenko's theorem allows for unbounded moments and represents a tail having a regular variation at infinity property that behaves like the fat-tailed pareto distribution (Feller, 1972).

Other approaches could be applied that follow a similar procedure with a GARCH filter such as filtered historical simulation (Barone-Adesi et al, 1998). However EVT is beneficial as it allows for low probability events that are out-of-sample, and more importantly, it allows for formal scaling across different holding periods. We could also scale GARCH volatility estimates directly but GARCH volatility for one horizon (due to time variation) may not give good estimates for multi-period horizons.

Our tail probability measure is obtained from the tail of the conditional distribution, F(z) for a single-period:

$$P_{Q,t} = (Z_{m,T}/Z_p)^{1/\gamma} m/T \qquad (4)$$

And the associated quantile measure is given by

$$Q_{P,t} = Z_{m,T} (m/Tp)^{\gamma} \qquad (5)$$

**Tail estimation**

In order to estimate the probability and quantile measures we need to estimate their key input, the tail measure, $\gamma$, for a given tail threshold, m, and we employ the commonly applied Hill (1975) semi-parametric tail estimator, $\gamma = 1/\alpha$, that operates analogously with EVT by dealing with order statistics. Beneficially this estimator describes the number of defined moments of a distribution and we need the variance to exist to allow the use of the EVT $\alpha$-root scaling law. Moreover, Kearns and Pagan (1997) find that the Hill estimator is the most efficient semi-parametric tail estimator when they compare it to both Picklands and de Haan and Resnick estimators. In contrast much of the literature on EVT follows parametric modelling by fitting a Generalised Pareto Distribution (GPD) to the data, but our approach only relies on the assumption that the data is fat-tailed as our financial returns series are, and thus use a semi-parametric estimator.

The Hill estimator has the same properties as the probability and quantile estimators and these are given in table 2. A weakness of the estimator is the lack of stability (often presented as a 'Hill horror-plot'). In essence estimation of the tail threshold, m, is non-trivial with potential small sample bias, and we use a number of approaches to ensure that stable estimates are obtained for the risk measures. We report values from using an *ad hoc* procedure of estimating the tail index for 1% and 5% of the data. We also present values using the modified Hill tail estimator, $\gamma_{hkkp}$, following Huisman et al. (2001) that minimises small sample bias and the impact of tail clusters. Huisman et al. (2001) find that their approach minimises overestimation of tail-fatness. The approach uses a weighted least squares regression of Hill estimates against associated numbers of tail estimates, $\gamma(m) = \beta_0 + \beta_1 + \varepsilon(m)$ for m = 1,….,$\eta$, and extrapolates the associated number of tail estimates, $m_{hkkp}$. We find the Hill values are reasonably stable regardless of approach used suggesting that the inferences from using any of these approaches to get probability and quantile estimators would be stable also. Moreover, regardless of the approach used the existence of a finite variance is confirmed with tail values in excess of 2 thereby supporting our use of the EVT scaling law.

INSERT TABLE 2 HERE

**Scaling procedure**

Thus far we have examined risk measurement for any (single) holding period. Our approach is extended for any time-frame using the EVT scaling law, known as the α-root of time. This scaling law acts analogously to the Gaussian square root of time scaling factor and does not require estimation of any additional parameters. Also there are efficiency gains in measuring the tail index at the highest frequency possible that minimises possible small sample bias.

Applying the scaling rule requires two conditions to hold: the data is iid and it exhibits a finite variance. Illustrating the scaling law we can adjust the asymptotic distribution of the fat-tailed Fréchet distribution by applying Feller's theorem (Feller, 1972, VIII.8):

$$P\left[\sum_{t=1}^{T} Z_t \leq z\right] = qF(z) \qquad (6)$$

Where q is the scaling factor (for $q = h^{1/\alpha}$). Our multi-period risk measures for any timeframe, h, assume that our risk measures in (4) and (5) are adjusted by the factor q. So our single-period probability measure in (4) becomes

$$P_{Q, h, t} = h^{1/\alpha} [(Z_{m, T}/Z_p)^{1/\gamma} m/T] \qquad (7)$$

And the related multi-period quantile in (5) becomes

$$Q_{P, h, t} = h^{1/\alpha} [Z_{m, T} (m/Tp)^{\gamma}] \qquad (8)$$

These probability and quantile estimators are then incorporated into (2) and (3) to give multi-period measures. We thus extend the conditional risk measures to any relevant holding period of interest.

It has been found (Cotter, 2007) that the EVT scaling law is more accurate than the Gaussian scaling law for conditional tail modelling. In particular, prior to scaling, there is an underestimation problem with assuming normality. Moreover, this underestimation becomes an overestimation problem when scaling to lower frequency. This is due to the fat-tailed distribution exhibiting a finite variance ($\alpha > 2$) and resulting in $\sqrt{h} > h^{1/\alpha}$. In a related paper McNeil and Frey (2000) use EVT to obtain the single-period estimates and they then use a Monte Carlo simulation to generate multi-period risk measures. Our approach is more efficient and easier to implement and exploits EVT to obtain high frequency tail estimates that are easily scaled for relatively low frequency tail estimates.

To summarise, our conditional EVT approach involves the following steps:
- Fit an AR-GARCH model to the returns series to get forecasts of the conditional mean and variance, and use the conditional volatility series to filter the returns resulting in iid residuals.
- Use EVT and obtain Hill tail estimates from the filtered iid series and associated single-period conditional risk measures.
- Scale the single-period conditional risk estimates by the EVT $\alpha$-root scaling law to obtain multi-period risk estimates.

**Simulation**

We now examine the properties of this approach through simulation and follow this with an application. We create a simulation for a sample size of 2000 with 200 replications of a GARCH (1, 1) model with parameters $\alpha_0 = 0.1$, $\alpha_1 = 0.15$, and $\beta_1 = 0.8$ underpinned by student-t innovations with 4 degrees of freedom. Thus the simulation encompasses volatility clusters and heavy-tails. We provide estimates of the probability and quantile measures for a single-period and for multi-periods (h = 2, 4 and 5) using the modified small sample hill estimator and the EVT $\alpha$-root scaling law.

Average estimates are given in table 3 where we present in-sample quantiles and out-of sample probability estimates. We also present the expected and actual number of violations for the quantiles to determine their accuracy. The precision of the findings is very favorable with the predicted values close to the true values both for single-period and multi-period settings. Moreover, the number of violations is close to what is expected. Generally the bias of the probability estimates is low when examining relatively low quantiles and this increases somewhat in moving to higher quantile losses. Furthermore, the multi period estimates see an increase in the bias in general from the single period estimates and this is most pronounced for the most extreme quantile threshold. Thus, the bias tends to result in an underestimation of the probability of experiencing very large losses and these are negatively related to the quantile estimates. Overall however it is important to stress that the bias is small.

INSERT TABLE 3 HERE

**An application**
Using the approach on daily S&P500 index returns, single-period and multi-period conditional probability and quantile risk estimates based on returns upto February 27, 2009, are given in table 4. To illustrate, the probability for any given day of having a negative return in excess of 5% for the S&P500 is 0.68%. This increases to a 1.04% likelihood over a weekly interval. Moving to the lower loss levels of 2% results in a higher probability estimate. Also, the multi-period forecasts have the advantage that the conditional environment is not measured at lower frequencies thereby avoiding losing the unique stylised features of relatively high frequency realisations and avoiding the dampening of volatility estimates. The scaled forecasts use iid returns as evidenced by the dependence structure of the S&P500 filtered series. Their accuracy is shown by the Monte Carlo study in table 3.[3]

Given the accuracy of the approach from the simulation study we also compare our scaled conditional estimates to that from assuming Gaussianity. We confirm that the single-period conditional normal estimates underestimate tail risk measures. Moreover this underestimation is reversed when we use the Gaussian $\sqrt{h}$ scaling in comparison to those from applying the α-root scaling law. For instance the probability of exceeding a return threshold of 2% is 8.84% (6.35%) for extreme value (Gaussian) values over a single day and this scales upwards to 13.58% (26.74%) for weekly intervals using the extreme value (Gaussian) scaling laws. As we have seen the simulation study suggests that the extreme value estimates are very accurate but suffer from a slight overestimation bias. Hence the multi-period estimates result in an overestimation bias that is far greater for the Gaussian estimates.

---

[3] The Monte Carlo results are very supportive of the approach. However we do not formally backtest the risk estimates for the S&P500 due to lack of data. For instance if we had a backtest using 1000 days we would need 5000 non-overlapping day returns to determine adequacy for the 5-day scaling rule.

INSERT TABLE 4 HERE

**Summary**

In summary we present conditional tail probability and quantile measures that account for fat-tails and volatility clustering. Investors can obtain risk measures that are updated for current mean and volatility values. Our conditional risk measures use an AR (1)- GARCH (1, 1) filter that results in iid returns and also allows for updating of conditional mean and volatility. We apply EVT methods to the iid filtered returns series using a modified Hill tail index estimate. The approach allows us to scale these risk measures for any holding period in an efficient and parsimonious manner using the EVT $\alpha$-root scaling law. We illustrate the benefits of the approach through a simulation study and application to futures data. In particular the approach illustrates the estimation bias that exists when assuming normality is minimised for the conditional EVT approach both for single-period and multi-period settings (using the respective scaling laws).

**Figure 1**

Plots of S&P500 Futures Contract

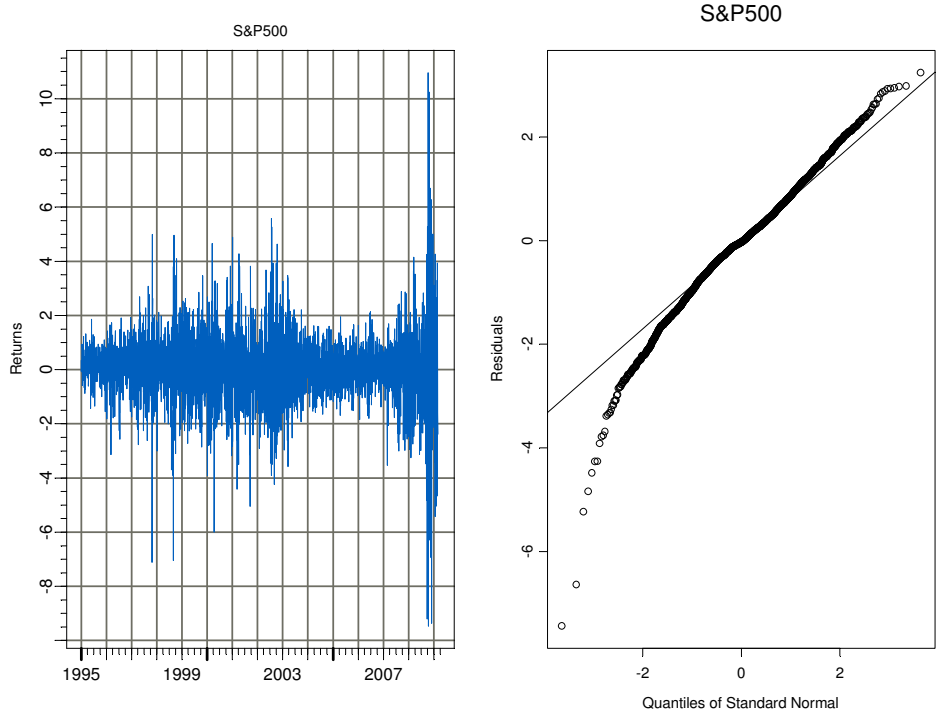

This figure shows the time series for returns and QQ-plot for the conditional distribution. The timeframe is January 1995 through February 2009.

**Table 1**
Conditional Modelling of S&P500 Index

| $\phi$ | $\alpha_0$ | $\alpha_1$ | $\beta_1$ | R(12) | $R^2$(12) | Z(12) | $Z^2$(12) |
|---|---|---|---|---|---|---|---|
| -0.030 | 0.006 | 0.066 | 0.924 | 56.723 | 3563.359 | 14.100 | 8.737 |
| (0.084) | (0.004) | (0.000) | (0.000) | [0.000] | [0.000] | [0.294] | [0.725] |

The AR (1) -GARCH (1, 1) specification assumes student-t innovations with 4 degrees of freedom. Marginal significance levels using Bollerslev-Wooldridge standard errors are displayed by parentheses. Serial correlation is examined using the Ljung-Box test on the returns (R), filtered (Z), squared returns ($R^2$), and squared filtered ($Z^2$) series. Marginal significance levels for the Ljung-Box tests given in brackets. * denotes significance at the 5% level.

**Table 2**
Downside Tail Estimates for AR(1)-GARCH(1, 1) Filtered S&P500 Index

| m1% | $\gamma$1% | m5% | $\gamma$5% | $m_{hkkp}$ | $\gamma_{hkkp}$ |
|---|---|---|---|---|---|
| 37 | 4.03 | 185 | 3.59 | 145 | 3.29 |
| | (0.66) | | (0.26) | | (0.27) |

Hill tail estimates, $\gamma$, are calculated for each futures index using the AR(1)-GARCH(1, 1) filtered returns. The threshold values, m1% and m5% relate to the one and five percentiles are used to calculate the associated tail estimates $\gamma$1% and $\gamma$5%. The number of values in the respective tails, $m_{hkkp}$, and the associated Hill estimates, $\gamma_{hkkp}$, follows Huisman et al. (2001). Standard errors are presented in parenthesis.

**Table 3**
Simulated GARCH (1, 1) with Student-t innovations and Scaling Procedure

|  | Single-period | Multi-period | | |
|---|---|---|---|---|
|  | h = 1 | h = 2 | h = 4 | H = 5 |
| **Probability** | | | | |
| **P25%** | 0.0019 | 0.0024 | 0.0032 | 0.0034 |
|  | (0.0020) | (0.0024) | (0.0028) | (0.0030) |
| **P50%** | 0.0003 | 0.0004 | 0.0005 | 0.0006) |
|  | (0.0005) | (0.0006) | (0.0007) | (0.0007) |
| **Quantile** | | | | |
| **Q95%** | 7.0413 | 9.1925 | 12.0010 | 13.0764 |
|  | (7.0900) | (8.4315) | (10.0268) | (10.6020) |
| **Q99%** | 13.0764 | 17.0714 | 22.2869 | 24.2842 |
|  | (13.6000) | (16.1732) | (19.2333) | (20.3367) |
| **No. Violations** | | | | |
| **Q95%** | 100 | 52 | 28 | 24 |
|  | (100) | (50) | (25) | (20) |
| **Q99%** | 24 | 13 | 4 | 4 |
|  | (20) | (10) | (5) | (4) |

The values in this table represent averages of 200 replications from a sample size of 2000. The blocks for the multi-periods used are h = 2, h = 4 and h = 5 loosely corresponding to 2 days, 4 days and weekly intervals. The probability and quantile estimates are based on Huisman et al (2001) tail estimates for the simulated data. The theoretical probability and quantile estimates are in parentheses. Values are expressed in percentages. The actual number of violations for each quantile is given with the expected number in parenthesis.

**Table 4**
Single-period and Multi-period Conditional Probability and Quantile Estimates for the S&P500 Index

|  | **Single-period** | | **Multi-period** | | | | | |
|---|---|---|---|---|---|---|---|---|
|  | h = 1 | | h = 2 | h = 4 | h = 5 | h = 2 | h = 4 | h = 5 |
| **Probability** | P5 | P2 |  | P5 |  |  | P2 |  |
|  | 0.68 | 8.84 | 0.82 | 0.98 | 1.04 | 10.63 | 12.79 | 13.58 |
|  | (0.00) | (6.35) | (0.12) | (1.64) | (2.81) | (12.21) | (23.12) | (26.74) |
| **Quantile** | Q95 | Q99.5 |  | Q95 |  |  | Q99.5 |  |
|  | 2.55 | 5.65 | 3.07 | 3.69 | 3.92 | 6.79 | 8.17 | 8.67 |
|  | (2.47) | (4.12) | (3.49) | (4.93) | (5.52) | (5.83) | (8.24) | (9.22) |

The values in this table represent the conditional probability and quantile estimates for different confidence intervals. For example, P5 is the probability of having a return exceed 5 percent and Q95 is the quantile at the 95% probability level. The estimates use Hill estimators based on the Huisman et al (2001) procedure from the AR(1)-GARCH(1, 1) filtered returns. The blocks of returns used are two days h = 2, four days h = 4 and five days (weekly) h = 5. Conditional estimates from fitting a GARCH (1, 1) model with normal innovations are in parentheses. Values are expressed in percentages.